\documentclass[11pt]{article}

\title{A Systematic Approach to Web-Application Development}
\author{Paul Fodor, Joy Dutta}
\date{}

\begin{document}
\maketitle
\tableofcontents

\newpage
\abstract
Designing a web-application from a specification involves a series
of well-planned and well-executed steps leading to the final product.
This often involves critical changes in design while testing the
application, which itself is slow and cumbersome. Traditional
approaches either fully automate the web-application development
process, or let developers write everything from scratch. Our
approach is based on a middle-ground, with precise control on
the workflow and usage of a set of custom-made software tools 
to automate a significant part of code generation.

\section{Introduction}
Web-application development is always a challenging task at hand.
With the ever changing internet technologies it
is difficult to master a particular technique and keep it
running without the risk of being too outdated and incompatible
with future technologies. While there are sophisticated 
frameworks for large scale web-application development, lack
of time-tested proofs of their reliability and maintainability
is still an issue with using them. Maintaining an existing 
piece of software is much more important because it is directly
related to the cost of it. Choosing a middle ground here proves
to be rational. A simpler framework with more code to
write from scratch can be helped by developing software tools 
to automate part of a large common code base which is 
otherwise error prone if written by hand. 

In our case, we have developed a medium scale web-application,
the Stony Brook University Graduate Application Web-app, using
entirely JSP/JavaBeans as the web technology.

While a sophisticated
framework like Struts or Spring can drastically reduce the amount
of code, the time to understand and master each of them can 
easily be inefficient for a medium scale project, supposed to be
maintained and monitored by different batches of students.

\section{Principles of task oriented Web design}

  \subsection{Choice of framework}
  Choosing a framework for a particular web-application is much
  like choosing the right tool for a job. Often a mix-up of current
  technologies prove to the right one for the task at hand. It
  needs significant background and future maintenance scenario
  analysis before coming to a concrete choice. The maintenance
  scenario is extremely important here because if an application
  is not going to be maintained by a fulltime trained 
  professional, a complex framework can easily spell disaster
  in the long run.

  One of the most stable and widely used web application
  frameworks is the JSP/JavaBeans/Servlets. Simplicity is
  one of its major strengths. There are many modern 
  frameworks widely used currently, like the Struts
  and the Spring frameworks, all of them employing the
  MVC model. There is also Hibernate, a database isolation 
  layer used with the latter two frameworks which offers
  few high level advantages over JDBC at the cost of initial
  setup and complexity overhead. 

  We have chosen our MVC model based on the JSP/JavaBeans/Servlets
  framework and resorted to JDBC for all communication with
  the Oracle database backend.
  
  \subsection{Design of page flows}
  Designing the flow of an application with primarily human
  interfacing (a web page with choice of operations) is inherently
  a task involving intuitive thinking. At the same time, the flow
  has to blend well with the backend where the data is processed.
  There is inherently a tradeoff when it comes to efficiency.
  In out case more emphasis is given to the intuitive page flow
  at the cost of adapting the backend to work with it. At the
  same time it has been kept simple by minimal use of session-saved
  attributes.
  
  An example is shown as follows. The scenario is of a new user
  attempting to start an application process. The user's new email
  and password must be setup before everything else, so a page
  with email and password information is presented to him/her.
  When the user fills the form, an email is sent to the given
  email address to verify its correctness. The email has a link
  which when clicked, confirms the user as authentic and is 
  allowed to proceed filling up other details in the subsequent
  pages.

\section{The Design of Graduate Application Webapp (GAW)}

  \subsection{Architecture}
  We made a generic architecture for the class of web 
  applications that concern online applications to universities.
  The advent of internet and e-commercialization of everything
  has lead the path to adhoc web application development.
  With a systemmatic development approach and architecture
  our goal is to make such application developments more 
  predictable and easier to build.

    \subsubsection{Users}
    We can divide the users of a graduate application system in
    a few major classes. At the bottom-most level, {\em prospective
    students} will be using it to apply to graduate school. 
    {\em Graduate school employees} can access the student records 
    and if needed, can modify some fields. {\em Reviewers}, assigned
    by the graduate school, who are normally professors in the university
    can review the resumes, transcripts and statements of purpose
    to comment on the students' chances of getting admitted.
    {\em Graduate directors} can assign graduate school employees
    for access on the student records. There can multiple graduate
    directors. At the top level, the {\em graduate application 
    administrator} can control everything about the system, 
    primarily assigning graduate directors.

    \subsubsection{Subsystems}
    The identifiable subsystems are:
    \begin{itemize}
    \item {\em Database interface} consisting of the methods to 
    communicate with the Oracle backend.
    \item {\em Client (user) interface} which is basically the jsp pages
    and makes the ''View'' part of the MVC model.
    \item {\em The beans} as the system ''Model'' with client input 
    validation logic.
    \item {\em The servlets} and {\em managers} as the ''Controller''
    of the system with the page flow logic.
    \end{itemize}

  \subsection{Application subsections}
  Here we discuss the form pages in the application process that an
  applicant has to fill up.

  \begin{enumerate}
    \item {\bf General Information:} Applicant's personal information
    like names, DoB, addresses and phone numbers are stored.
    
    \item {\bf Application Information:} The academic program, semester
    and attendance status is stored.
    
    \item {\bf Educational History:} Applicant's previous education
    records such as college names, addesses, subjects, degrees and
    GPAs are stored.
    
    \item {\bf Employment History:} Applicant's previous employment
    records such as company names, addesses, positions and work dates
    are stored.

    \item {\bf Qualifications:} Additional credentials like journal
    articles, conference papers, project reports are entered here.
    
    \item {\bf Test Scores:} Test and score information of a varied 
    number of tests (e.g., GRE/TOEFL) that the applicant had taken
    are stored.

    \item {\bf Language Proficiency:} Information about proficiency in
    english of international applicants are entered.
    
    \item {\bf Financial Aid:} Applicants report if they received 
    fellowships like AGEP, EOP, SEEK etc.
    
    \item {\bf Resume and Statement of Purpose:} Applicants can either 
    upload their resume/sop in any well-known format (.txt, .doc, .pdf etc)
    or can type it in the input area provided.
    
    \item {\bf Recommendations:} Information about recommenders are 
    put by the applicant. Later when the recommenders fill up the 
    recommendation text they also enter comments on the applicant's
    academic performance and motivation and the chance of his/her being
    a potential TA in the department.
    
    \item {\bf Transcripts:} Applicants can either 
    upload their transcript in any well-known format (.txt, .doc, .pdf etc)
    or can type it in the input area provided.

    \item {\bf Supplemental Department Applications:} Additional 
    application information that are unique to each department are 
    entered here.
    
    \item {\bf Check Applications:} Applicants can check for validity of
    their whole application process before submitting.
    
    \item {\bf Submit and Pay:} Applicants can pay using an international
    credit/debit card. The payment is done securely by using the service
    verisign which is a proven portal for secure third party payments.

  \end{enumerate}  

  \subsection{Code hierarchy}
  To effectively handle a large code base, we have partitioned
  the java code in manageable concepts or entities other
  than the jsp pages, namely {\em bean}, {\em manager},
  {\em servlet} and {\em util} which are also the package names. 
  The top level package is {\em gaw}. The package descriptions
  are briefly given below:

    \subsubsection{gaw.bean}
    The bean classes correspond the tables in the database. The 
    variables in a bean are one-to-one with the fields in the 
    corresponding table. They basically act as containers for 
    one row of data to be moved around the application. The beans
    are used in the jsp pages to populate the fields and to 
    store data around session and are saved as session attributes
    for the duration of the active session.

    \subsubsection{gaw.manager}
    The manager classes are designed as a layer between the beans
    and database, to perform the CRUD operations. In more complex
    frameworks this is where we have Hibernate, and much more 
    powerful abstraction layer. Using the managers we retrieve
    a variety of table information, either one row corresponding
    to a primary key, or a set of rows satisfying some criteria.
    
    \subsubsection{gaw.servlet}
    Servlets are part of the web-application framework we are using.
    These take the http request coming from the forms in the jsp
    pages and control the page flow. They set the session attributes
    with the updated beans, call the managers to save or update the
    data and redirect to the next appropriate page. It also calls
    the validation routines in a bean before saving it and if the
    request page form has errors in one or more fields, the same
    page is redirected to again with the error list at the top of it.
    
    \subsubsection{gaw.utils}
    A set of utility classes are made, namely the 
    {\em DBConnectionPool},
    {\em EmailHelper} and {\em ConfigReader} to provide some of the
    basic core functionality. 
  
    The {\em DBConnectionPool} class provides the pooling of database
    connections. Currently a singleton JDBC connection object is used,
    but with increasing load in future, multiple connections can be 
    kept open and load-balanced the use between them.

    The {\em EmailHelper} class provides e-mail functionality with 
    just a function call. The configurations for the email are 
    externally specified in a configuration file. It is used to send
    mail to a new applicant to verify authenticity of his provided
    email to log in and use the system.

    The {\em ConfigReader} class is essentially a tool to read a 
    xml configuration file for various configurations of the system.

  \subsection{Functional specifications}
  In this section several key functionality of the system will be 
  discussed, in the use-case level. The idea of intuitive page flows
  govern the majority of functional design. Many of the design 
  decisions stem from common sense only.

  \begin{itemize}

  \item When a new prospective student applies for a graduate program,
  he is first required to input email address and password. A confirmation
  mail is then sent to the email address to verify its authenticity. when
  the user clicks the link in that mail, he becomes a confirmed user
  and can log in to the system to start the application process.
  
  \item An user can change his/her email address. Even though email 
  should be unique, it is not a primary key because it is changeable.
  the primary key for a user is user\_id, which once created, doesnt 
  change ever.
  
  \item All the jsp pages check the existence of a user email address
  in a session attribute, as they are all contextual, needing a valid
  user for the page content to be shown. If a session is not on, a
  message is shown requesting to log in first.
  
  \item If an applicant deletes his/her account, all the information
  about the account is still retained, only the email address is 
  prefixed with a special word so that it can no longer be used.
  
  \end{itemize}

\section{Implementation}
In this section several key approaches towards a systemmatic
software development is discussed, focussed to web-applications.
Naming fields, variables and methods is one of the most
important task, given the finite short term memory capacity
of human developers. This is closely followed by some custom
software tools whenever code can be churned out mechanically,
following a rule.
  
  \subsection{Design of the database}
  The database design follows the principle of generalization 
  of all possible applications for graduate school. Besides 
  having a separate table for login information which contains 
  the email address and password of the user, this table also 
  contains the user\_id field. The user\_id field is the
  unique identifier of the user in the whole system. This id is 
  invisible to the user and cannot be updated of changed. In this 
  way the user can change his login informations and email, 
  operation which does not require any change in other database 
  tables. Majority of tables in out design contains a primary 
  key in the form of an id, which is auto generated, used to 
  refer a row from other tables.
  
  \subsection{Naming}
  Naming variables is one of the most daunting tasks in any
  large scale software development. In most cases involving database
  tables, variables and methods closely depend on the fields in the
  tables. If the naming of variables is not done systemmatically,
  it is easy to make errors when dealing with tables, fields, variables
  and method names. In a web-application involving JSP/JavaBeans
  technology, a big component of the code is the beans where we
  practically have a one-to-one mapping of variables with the fields
  in the database tables.
  
  In a web-application, many things depend on the field and table
  names in a database. Table names dictate bean class names and 
  fields dictate bean variables, getter and setter method names.
  
  A standard naming approach simplifies coding and recalling a 
  number of variables and method names instantly from a field or
  table name. We used a formal naming approach as follows: if a 
  database field is {\tt first\_name}, the corresponding
  variable name would be {\tt firstName} and the getter/setter
  methods would be {\tt getFirstName()} and {\tt setFirstName()}
  respectively.

  \subsection{Code generation}
  A major part in a web-application based on JSP/JavaBeans is
  the development of the bean and the manager classes. 
  A bean consists of a
  set of varibles, usually corresponding to a set of fields
  in a database table and their respective getter and setter
  methods. There are also other added aspects like error
  checking approach embedded in the beans. We discuss it in a 
  later subsection.

  A manager is a controller class which takes care of the
  CRUD (create, retrieve, update, delete) operations of the beans
  with the database.

  Coding of managers is a time-consuming and repetitive process
  and often prone to typographical errors. Errors with closely
  similar variable names are hard to detect and debug as
  they do not break compilation, so there must be some sort of 
  automation to generate all the common code. 

  A tool called BeanHelper is developed for this particular task
  at hand. It helps with fragments of code concerning putting the
  bean variables in the database using a sql statement and 
  retrieving them. We discuss more about it later.
    
  \subsection{MVC model}
  We employed a Model-View-Controller model on our approach.
  The main code is broken up in the following parts: jsp, beans, 
  managers and servlets. Obviously the jsp pages constitute the
  "View" in the MVC model as they are interface of the system with
  the outer world. The beans hold the data and the logic for its
  validation, hence they are the "Model". The managers basically
  take care of the CRUD (Create-Retrieve-Update-Delete) interface
  of the user data with the database using the beans as data
  containers. The servlets receive the user requests when forms
  from the jsp pages are submitted and process them, using managers
  to interact with the database and control the session. So the
  managers and servlets constitute the "Controller".

  \subsection{Database interface}
  The quickest way to interface a database is using JDBC with
  a couple parameters. On the other hand in frameworks like
  Hibernate, there is an extensive tiering of configurable
  layers. In our system we made use of JDBC with parameters
  set using a simple XML configuration file.

  We employed the concept of connection pooling to make use
  of load balancing later on when there are simultaneous
  activity and significant speed-up is achievable by several
  parallel connections to the database. At present only one
  connection object is created and is re-used between 
  database connection open requests.
  
  While saving a row of data from a bean to a table, the 
  method in the manager class is called {\tt saveOrUpdate()}
  because we have no information beforehand if the data
  is new or an update of an existing row in the table.
  We check for the primary key for existence and depending
  on the outcome do either an insert or an update operation.

  \subsection{File uploads}
  For uploading files like resumes, statements of purposes
  and transcripts, we used a third party file upload library
  {\em Upload Bean}\cite{uploadbean}.
  
  Instead of writing code for modifying each individual tables 
  about the upload status, this library helps reducing code 
  duplication in a novel way. It uploads any file to a special 
  table and provides us a key for the upload, i.e., an 
  upload\_id. We only put that upload\_id in any application table
  which concerns the uploaded file. This made us write 
  once one chunk of code for the upload, with the destination
  table information as the parameters.
  
  \subsection{Error validation with Java beans isValid methods}
  Checking for validity of every field in a form is one of the
  biggest drudgery in the coding of web-applications. This
  is also the often ignored part in the process but one of the
  most critical as well. Security errors can easily creep in
  in the form of SQL injections and other attacks if the fields
  are unchecked before processing the contents. There are tedious
  frameworks with formalized verification laguages which can
  be in the jsp page itself, with custom tags, but this is out
  of our approach. In our method, instead of verifying the 
  correctness of the input data in the "View", i.e., jsp pages,
  we put the verification logic in the "Model", i.e., the beans.

  By default, a field is validated by checking if it is empty.
  More complex validations include date and country-state 
  combinations. Tools are made to validate if a given date 
  is after a configured "oldest possible" date or before the
  "newest possible" date. In country and state validations,
  it must be checked that "USA" as country choice is associated
  with a US-state from a drop down list and any other state
  is accompanied by a non-USA country.
    
    \subsubsection{AJAX approach and its limitations}
    A significant part of today's web applications are increasingly
    using a new technology - Asynchronous Javascript and XML (AJAX).
    We have been so far only familiar with the transfer of a whole
    web page from the server to a client, even in the event of 
    a small change of data coming back to the client. This produces
    a significant delay in the feedback typically when client
    submits a form and the server sends the form back with 
    fields containing invalid data in a different format, like in 
    red color. Using AJAX this wasted time and bandwidth can be 
    entirely eliminated.
    
    However, all good things come with at least some drawbacks.
    Using AJAX without a design in mind introduces a problem
    with browser state. Simply put, the back button of the 
    browser will no more retain its use in AJAX-enabled 
    pages. For example, if a page with an AJAX interface is
    visited and the data on that page is refreshed several 
    times by the AJAX functionality, the back button will
    take to you to the previous page instead of going through
    the previous data on the same page. Since the back button
    is fairly intuitively used by users, care must be taken 
    to present a highly visible, alternative solution to the
    back button functionality, preferably in the form of links
    that call some javascript function to help retrieving the
    previous state of the same page.

    AJAX applications are also difficult to debug because the 
    processing logic is embedded both in the client and on the server.
    The client-side JavaScript code may be viewed simply by 
    selecting View Source from an AJAX-enabled HTML page. A poorly 
    designed AJAX-based application could open itself up to hackers 
    or plagiarism.

  \subsection{Error and Exception management}
  A significant part of the design process of any large software
  is the handling of errors and exceptions. This section alone 
  deserves a detailed article covering every nuances of its use.
  Theoretically, the logic flow in any large process will work
  correctly for a very small set of data that are ''valid''.
  Any one ''invalid'' data has the potential of disrupting a
  whole chain of linked processes and workflows. The way 
  exceptions, warnings, errors and fatal errors are handled
  goes a long way towards distinguishing a great software
  from a crappy one.
  
  From my experience, exceptions are best handled in as much 
  far in the top level as possible. An user applying for
  a graduate program may not make anything out of a file
  open error that is deep inside the system, but a more
  generic message saying the system has encountered some
  problem will make much more sense.

  A lot of methods in Java's extensive API throw a variety of
  Exceptions, all of them are subclasses of the most generic
  class {\tt Exception}. A few guidelines can lead to a better 
  handling of exceptional cases:
  \begin{enumerate}
  
  \item {\bf Exception forwarding.} In a series of nested method calls 
  it is more logical to forward an
  exception to the caller if the callee can't handle or repair the case.
  Suppose a function is supposed to open a config file and return its 
  contents. If for some reason it can't find the file, it will get a 
  FileNotFoundException. Now it can do two things:
  It can return an empty string if the caller doesnt bother to
  know if any error happened. Or it can forward the exception to the 
  caller by throwing it explicitly.
  
  \item {\bf Exception wrapping.} Like exception forwarding but 
  a number of specific exceptions (subclasses of {\tt Exception})
  can be wrapped up in a application specific custom Exception class.
  For example, instead of throwing a FileNotFoundException or
  IOException to a caller method, it can be wrapped in a custom
  exception class GawConfigException and thrown to the caller.

  \end{enumerate}
    
  \subsection{Logging}
  The common tendency for developers to use console output
  for debugging as well as feedback can pretty much be termed as
  "bad habits" in software development. Using printf, cout or
  System.out.println() can quickly be messy and actually
  an obstacle for reading an otherwise well written code.
  One added fact is that when developers try to debug a part
  of the code, they put arbitrary console output statements
  that must be removed or commented after the debug is 
  finished. There are quick ways to do this type of enabling
  and disabling of debug statements, but they are inefficient
  at best for use in a large scale software.

  An industry-wide used solution for this task is logging.
  Logging involves using an API designed for a multi-tiered
  approach to outputting a varied levels of messages to
  a number of configurable destinations. Using the logging
  infrastructure, turning debug messages on/off alongwith
  all kinds of change of preferences can be done using a
  configuration file. Moreover, message output destinations
  are pluggable, this means we can add a file, a database
  connection, or even an e-mail to the message output stream.
  The e-mail plugin is helpful when we are interested in
  being notified for some critical/fatal errors happening
  in the system during runtime.
  
    \subsubsection{log4j}
    One of the most useful java packages is the log4j API for
    logging. The startup being a single line at the beginning 
    of a java class and a simple configuration file in the 
    current directory, there is no reason to not use it
    for the benefits it introduces to software development of
    any kind.
    
    log4j employs several key concepts: The message severity
    levels or {\em Priorities}, The pluggable message destinations
    or {\em Appenders} and the hierarchical logging sources
    or {\em Categories}. The log messages can be formatted
    by a format string. Everything above can be configured using
    a config file.

    {\em Priorities} are several levels of messages, like
    DEBUG, INFO, ERROR, WARN, FATAL in the increasing order
    of severity. The basic idea is that a threshold can be set
    so we can only be interested on messages above a specified
    priority.
    
    {\em Appenders} are pluggable destinations. A message can be
    channeled to multiple destinations that are interested to
    be notified. For example, we can plug console as well as 
    a file for a particular category of messages. Each appender
    can set its own threshold of messages, so we can set the 
    console to show all messages upto DEBUG, while the file
    only will store messages upto the severity INFO. Each
    appender can have a different output format too.
    
    {\em Categories} are like full java class names with package
    names. Basically a dot separated notion to employ the idea
    of inheritance. A category {\tt foo.bar} will inherit properties
    from its parent category {\tt foo}. This is particularly useful
    when used inside code base that is divided into multiple 
    hierarchies, like we did.
     
  \subsection{Tools}
  A number of assistive software tools have been made to help
  providing a library of useful methods like making connection
  to database, sending an email and generating code for some parts 
  of the system. Once properly
  tested, this provides immunity from common typographical
  errors that can potentially slow down the dev-test cycle.
    
    \subsubsection{EmailHelper}
    This is basically an email tool to send out email. During
    out dev-test cycle we have used the free POP/SMTP service 
    of softhome.net till now. Currently we use a gmail account
    as the previous one stopped their free service. To use the
    SMTP service of gmail, other than using the port 465,
    we needed to specify a few more properties to enable sending 
    mail using TLS. They are as follows:
    \begin{verbatim}
    props.put("mail.smtp.starttls.enable", "true");
    props.put("mail.smtp.socketFactory.port", "465");
    props.put("mail.smtp.socketFactory.class", "javax.net.ssl.SSLSocketFactory");
    props.put("mail.smtp.socketFactory.fallback", "false");
    \end{verbatim}
    
    \subsubsection{BeanHelper}
    While we wanted the manager classes would be generated
    from specifications, it was going to involve a rigorous tool
    that would have taken a lot of time. All the managers
    use a fair amount of code that involve getting/setting
    bean variables, retrieving parameters from http request
    and so on.

    The BeanHelper tool outputs a series of code fragments that
    are supposed to be copy-pasted in various parts of the
    manager code. The output consists of multiple formats
    that concerns different parts of the functions in the
    manager class.
    
    \subsubsection{TableConv}
    It is always desirable to have an isolated testing environment
    of a web-application, including a separate database space.
    The real database where the data from a production version
    of the web application will actually go is an important one
    and can't be offered as a testing ground over the duration
    of the development. This necessitates a tool which can operate
    as a bridge between the application-specific database and the
    global database and can be used in a caching-like mechanism.
    Whenever there is a pool of new data collected in a session
    with the web application, the tool can dump them together to 
    the global database in one go.

    The table converter tool basically maps fields from one
    table to another, and it can be configured such that inter
    database table mapping becomes possible.

\section{Efficiency}
  
  \subsection{Code redundancy reduction}
  Code redundancy is by far the most unwanted and ill aspect
  in any large code base. Often developers opt for a quick
  fix at some part of code by borrowing code from another part
  which is essentially a replica of a code fragment. This
  kind of quick code-fix is nothing but a vulnerable place
  for a bug if one of those two replicas is later found to
  contain a bug and is fixed while the other place is forgotten.
  
  To handle this problem, code re-use is one of the most 
  important approaches according to best practices of software
  development. A replicable code can be made as a function
  and placed at a convenient place like a library and can
  be called from as many places without the risk of having
  stale code. Obviously re-use is a valid way to deal with
  the situation where we need exactly same code in more than
  one place, but doesn't concern using similar but not
  identical codes. In most cases we face the problem of 
  having to copy a chunk of code from another place and change
  a few variable names, leaving the logic same as before.
  This is another significant form of avoidable redundancy,
  because if the logic changes later in one place it has to
  be changed manually in other places with different variables.
  
  In our approach we attack the problem of similar code
  redundancy by generating code that are functionally similar.
  As a concrete example, we are generating the java beans
  which are basically a container with a set of variables and
  getter/setter methods for those variables. Once we know the
  bean class name and the variable names, the rest is pretty
  much mechanical, except a few places to fine tune.
  Generation of those beans from a simple specification and 
  a custom tool saved a lot of time hand-coding them. Furthermore,
  since the beans depend on the database tables, with each minor
  change in a table the regeneration of a compilable bean class
  becomes a breeze.

  We also discuss another probable speedup in our web-application
  development, after spending hours on hand-coding the jsp pages.
  As we discovered an approximate pattern, though much more complex
  than the beans, a specification for a jsp form page can be 
  written and the forms generated much like the beans with our 
  ability to fine tune for the looks and other stuff related to the
  "View" component of the MVC model.
  
    \subsubsection{Form generation - an approach} 
    A form page can be simply viewed as an interface to insert or update
    a table row or a part of it. In most cases, the fields in a form
    exactly corresponds to the fields in a database table. Besides
    the jsp form pages are basically a function of a few parameters
    like the table name, fields, the action servlet and so on.

    It would be a logical approach to generate form pages based on
    a table schema. The key challenges involved are the types of
    the fields and the proper interfaces to present the data. A typical
    example of this is the choice of using radio buttons or a checkbox
    for a boolean field. This task can be best handled by a well
    formed specification language and a custom tool to generate a 
    form from the specification.
    
    \subsubsection{AppFuse}
    AppFuse is a very radical approach to web-application development
    by Matt Raible. Given a database schema, it generates
    the bulk of the common CRUD (Create, Retrieve, Update, Delete) 
    code, which can be customized for individual use. It 
    extensively uses new technologies in every tier, and a 
    drawback of using it is the need of in-depth knowledge of
    all the components, mainly the Spring/Struts MVC model and
    the Hibernate database abstraction layer.
    
    Appfuse has a great potential to help with this scale of 
    web-application development, but needs intensive training to 
    master the fairly complex frameworks.
    
\section{Security issues}
Like many standalone softwares are prone to attacks like buffer 
overflows and format string exploitations, web applications
have their own set of vulnerabilities. In this section we 
will discuss a few of them and how much protection our system
has from them.

  \subsection{SQL injection}
  This is a common vulnerability of a web-application. When
  the server side code communicated with a database by an SQL
  query using a parameter sent by client in a query string,
  SQL injection is possible. For example, if the client is
  supposed to send a "name" parameter to server and the server
  makes the following query:\\\\
  {\tt "SELECT * from login where name = " + name}\\\\
  where the variable {\tt name} will have the client supplied data.
  An injected malicious data can be something like:\\\\
  {\tt 'john'; UPDATE login SET root\_access = 'Y' WHERE 
  name = 'john'}\\\\
  This will actually make two statements which upon execution 
  give some authoritative access to the user john.\\
  
  \noindent
  However, this injection works in php as a research paper 
  \cite {taint} suggests, but as soon as the java code 
  sees the semicolon separating two SQL statements, it throws
  up an exception. JDBC doesnt allow executing two statements
  like this, hence our system is secure from this exploit.

  \subsection{Directory traversal}
  In a directory traversal attack, an attacker attempts to access
  files outside of an authorized directory, e.g., the document
  root in the case of a web server. This is usually done by 
  including ".." to ascend above the document root. If there is a 
  check for "..", they can use hexadecimal representations of the
  characters.

  However this attack is more suitable on a ftp server than a
  tomcat server that we are using. Tomcat wont let us using ".."
  maliciously and it can only result in showing upto Tomcat's
  home page.

  \subsection{XSS: Cross Site Scripting}
  Cross site scripting (also known as XSS) occurs when a 
  web application gathers malicious data from a user. The data is 
  usually gathered in the form of a hyperlink which contains 
  malicious content within it. Usually the malicious portion is
  encoded in hex so it looks less suspicious.

  Common places to find XSS are large websites and bulletin boards.
  Some XSS can automatically execute when opening email or attachments
  or even just reading a public guestbook or forum post. One of the
  best ways to avoid any harm from these is to turn off javascript.

  \subsection{Does SSL actually offer protection ?}
  Websites that use SSL (https) are in no way more protected 
  than websites that are not encrypted. The web applications work 
  the same way as before, except the attack is taking place in an 
  encrypted connection. People often think that because they see 
  the lock on their browser it means everything is secure. 
  This just isn't the case.

  \subsection{Secure payment using Verisign}
  We used Verisign, a proven, secure third party online payment 
  system for the credit card payments. We pass the userId and when 
  the transaction is approved we do a silent post and after the 
  submit we return to the original page. In this way we don't 
  lose any transaction even if the user closes the window when 
  he gets the receipt from verisign. Since at the beginning of the 
  transaction we pass the userID, on the return from the 
  transaction we save the activation number for the user. In this 
  way the processing of the credit card is passed to a secure 
  specialized website and there won't be any information about 
  the credit card saved on our side.

\section{Special security issue: XSS}

The cross-site scripting attack is one of the most common yet
overlooked problems that web developers face today.
Cross-site scripting works by embedding malicious code on web 
pages with tiny "scripting" programs usually embedded in a 
hyperlink. When an unsuspecting visitor clicks that link it 
activates the hacker's program by using the corrupted script. 
Once activated, the rogue program allows the hacker to slip 
undetected past firewalls to read steal information from cookies, 
credit card numbers and other data.

  \subsection{Who is at risk}

  Mainly dynamic web sites are at risk from this type of attacks. 
  Static sites should be free of concerns. Dynamic web sites like 
  online forums face this risk because of their capability to 
  generate pages on the fly based on unvalidated input, like when 
  some random user posts a hyperlink crafter with malicious script 
  content.

  Even having an SSL-enabled website is not a cure from XSS. The 
  attacks work the same way as before, only in an encrypted 
  connection in the case of SSL.

  Any poorly coded script is a potential target because they 
  sit on the boundary of outside data and the server and usually 
  has the ability to execute commands powerful enough to do damage 
  to the system. If the script isn't careful about what data it is 
  processing it could be hijacked by some clever scripting data, 
  leading to disaster.

  \subsection{Mechanism}

  Most web browsers can interpret scripts embedded in web pages. 
  For example, with <iframe> tags, a script will be executed upon just 
  viewing the page. Such scripts may be written 
  in a variety of scripting languages and are run by the client's browser.
  Most browsers are installed to run scripts enabled by default.

  The malicious code embedded in a hyperlink can be encoded as hex
  to make it look less suspicious.

  \subsubsection*{Malicious code provided by one client for another client}

  Sites that host discussion groups with web interfaces have long 
  guarded against a vulnerability where one user embeds malicious HTML 
  tags in a message intended for another user. For example, an attacker 
  might post a message like

  \begin{verbatim}
      Hello. Start of message.
      <SCRIPT>malicious code</SCRIPT>
      End of message.
  \end{verbatim}    

  When a victim with scripts enabled in their browser reads this message, 
  the malicious code may be executed unexpectedly. Scripting tags that 
  can be embedded like $<$SCRIPT$>$, $<$OBJECT$>$, $<$APPLET$>$, and $<$EMBED$>$.

  Most discussion group servers either will not accept such input or 
  will encode/filter it before sending anything to other users.

  \subsubsection*{Malicious code sent inadvertently by a client for itself}

  A situation may occur when the client relies on an untrusted source of 
  information when submitting a request. For example, an attacker 
  may construct a malicious link such as

  \begin{verbatim}
  <A HREF="http://example.com/comment.cgi?mycomment=<SCRIPT>malicious code</SCRIPT>"> Click here</A> 
  \end{verbatim}

  When an unsuspecting user clicks on this link, the URL sent to example.com 
  includes the malicious code. If the web server sends a page back to the 
  user including the value of mycomment, the malicious code may be 
  executed unexpectedly on the user's computer. This example also applies 
  to untrusted links followed in email or newsgroup messages. 

  A simple example of cookie theft is given below:
  \begin{verbatim}
  http://www.example.com/search.pl?text=<script>alert(document.cookie)</script>
  \end{verbatim}

  If an attacker can get us to select a link like this, and the web 
  application does not validate input, then our browser will pop up an 
  alert showing our current set of cookies. This particular example is 
  harmless but an attacker can do much more damage, including stealing 
  passwords, resetting home page, or redirecting to another website.

  Even worse is the case of $<$IFRAME$>$ tags. Scripts inside that can
  run by just viewing the page.

  XSS holes allow Javascript injection, which may allow limited 
  command execution. If there are browser holes as well, commands
  can be executed on the client's side as well. So XSS holes can be 
  used to possibly exploit browser holes as well.

  \subsection{Ways to prevent}

  The easiest way of protection is to avoid clicking on links posted by
  arbitrary users on a web forum. Sometimes XSS can execute automatically 
  upon opening email or reading a forum post. Some caution can be exercised
  in these cases by turning off Javascript in the browser settings. This
  can prevent cookie theft too.

  One of the ways to thwart XSS is to sanitize the input data which
  comes when some innocent user clicks a carefully crafted hyperlink made
  by a hacker. For example this data can come through the \$QUERY\_STRING
  variable in the context of CGI programs.
  Most of the time the culprit are $<$script$>$ tags and javascript function 
  calls, so removing the characters '$<$', '$>$', '(', ')' should be a good
  first step. But, instead of removing bad characters, a better and 
  recommended approach is to define a list of acceptable characters
  and replace any character not in the list by an underscore. In that way
  the programmer becomes certain that whatever string is returned, it
  contains only characters under his/her control.

  A basic perl code for this should look like:

  \begin{verbatim}
  #!/usr/local/bin/perl
  $_ = $user_data = $ENV{'QUERY_STRING'};	# Get the data
  print "$user_data\n";
  $OK_CHARS='-a-zA-Z0-9_.@';	# A restrictive list, which
				  # should be modified to match
				  # an appropriate RFC, for example.
  s/[^$OK_CHARS]/_/go;
  $user_data = $_;
  print "$user_data\n";
  exit(0);
  \end{verbatim}

  In our web application, special checking for every user input
  needs to be done. We use POST mechanism to submit form information
  to server. In the backend, the servlets are coded to make any GET 
  request also handled the same way as POST. We can disable any use of 
  GET mechanism, but still some attacker might craft a malicious
  script in one of the textfields as user data when filling up a form.
  As a possible solution, during the field validations, all the fields
  must go through a sanity check function before other data validation
  should take place. A warning can be issued if the user data contains
  unacceptable characters like mentioned in the previous paragraph.

\section{Concurrency issues}
{\em Concurrency} refers to the sharing of resources by multiple 
interactive users or application programs at the same time. 
When developing such an application, care should be taken to prevent 
undesirable effects, such as:

\begin{itemize}

\item {\bf Lost updates.} Two applications, A and B, might both read 
the same row from the database and both calculate new values for one 
of its columns based on the data these applications read. If A updates 
the row with its new value and B then also updates the row, the update 
performed by A is lost.

\item {\bf Access to uncommitted data.} Application A might update a 
value in the database, and application B might read that value before 
it was committed. Then, if the value of A is not later committed, but 
backed out, the calculations performed by B are based on uncommitted 
(and presumably invalid) data.

\item {\bf Non-repeatable reads.} Some applications involve the following 
sequence of events: application A reads a row from the database, then goes 
on to process other SQL requests. Meanwhile, application B either modifies 
or deletes the row and commits the change. Later, if application A 
attempts to read the original row again, it receives the modified row 
or discovers that the original row has been deleted.

\item {\bf Phantom reads.} The phantom read phenomenon occurs when:
  \begin{enumerate}
    \item Application A executes a query.
    \item Another application B inserts or updates data that 
    satisfies A's query criteria.
    \item Application A repeats the query from step 1 (within the same 
    unit of work), but the result set is different because it includes 
    additional "phantom" rows inserted or updated by application B.
  \end{enumerate}

\end{itemize}

Such concurrency issues can be prevented in the application by managing 
locks and isolation levels. If the application does not require multiple 
database connections, then we can avoid concurrency issues altogether 
by disabling shared access. For example, the {\tt connect()} method in the 
java.sql.Driver interface supports ENABLE\_SHARED\_DATABASE\_ACCESS, 
a boolean property that can be set to false to disable concurrent access.

Since our application is going to handle sufficiently low volume of 
user interactions, we can safely turn off the above key. The 
ENABLE\_SHARED\_DATABASE\_ACCESS key is set to false by default
hence our system is free from the concurrency issues.

\section{Performance}
After developing any software it is customary to evaluate its
performance. Often a standardized method called benchmarking
is used to test a software and also to compare its performance
with other similar products in market. Regression tests measure
the worst case performance.

A web application is unlike a standalone software in the area
of testing. Test suites can be written for a standalone
software to quickly and repetitively test individual methods 
or a sequence of operations expected in normal use.
Web applications require client inputs through interactions
in web pages and is much slower to test. For every test case
the client has to click a series of buttons and fill up forms
in a particular order.

A basic UI performance test is the measure of response time
of form actions within an acceptable time frame. Table
\ref{uiresp} summarizes the acceptable response times for any
UI application.

\begin{table}
\centering
\begin{tabular}{lp{4in}}
\hline
{\bf 0.1 second} & is about the limit for having the user feel 
that the system is reacting instantaneously, meaning that no 
special feedback is necessary except to display the result. \\
{\bf 1.0 second} & is about the limit for the user's flow of
thought to stay uninterrupted, even though the user will notice 
the delay. Normally, no special feedback is necessary during 
delays of more than 0.1 but less than 1.0 second, but the user
does lose the feeling of operating directly on the data. \\
{\bf 10.0 seconds} & is about the limit for keeping the user's
attention focused on the dialogue. For longer delays, users will
want to perform other tasks while waiting for the computer to
finish, so they should be given feedback indicating when the
computer expects to be done. Feedback during the delay is
especially important if the response time is likely to be highly
variable, since users will then not know what to expect. \\
\hline
\end{tabular}
\caption{Acceptable UI response times \label{uiresp}}
\end{table}

Even though our system is not tested on a production machine,
it performed pretty much according to the above table.
The worst case time it takes for a page to load is a few seconds
when it loads a drop down list of all academic programs available
in the university, which is pretty huge.

Due to time and resource crunch, we could not perform regression
test on the system. But since this is not a web application
which many users will use often, simultaneously and repetitively
like a typical financial or dating portal, we can estimate that
our system should behave normally under normal use and 
circumstances.

\section{Future improvements}
Like in any software development process, a few things were learnt
in the later stages of growth of our web application. Lack of
time and resources prevented us from getting the best out of the
learning, but if properly documented, it might help someone to
easily implement it in future.

  \subsection{Role-based access control (RBAC)}
  Initially our system started with only the student applicants 
  in mind. Our resources were spent making a well designed
  web application but with little focus on the future growth in
  terms of different classes of users and their varied forms
  of accesses. As the application is going to deal with 
  personal and possibly confidential data about prospective
  students, it is highly important to have a proper access 
  control mechanism in place.
  
  Role-based access control seems to be a fair candidate in this 
  situation. A multitude of roles can be configured and tuned 
  for the right access control. Users can just be assigned a role, 
  and when the access for a role needs to change, the necessary
  access modifications can take place in only one location, instead
  of all the affected users.
  
  \subsection{Form field validation using custom tags}
  Even our systemmatic approach towards server side validation 
  using the beans as container of validation logic seems laborious.
  In the frameworks like struts, custom tag libraries for common
  field validations can be used for much easier development 
  workflow.  Custom validation logic can be specified in an 
  XML document, removing the necessity to recompile java code for 
  every modification.

\section{Using CVS and parallel development}
For projects of this scale, version control should be a mandatory
practice. Not only does it ensure good collaboration between multiple
developers, having the repository residing on a stable servers removes
a lot of vulnerability of carrying the code in personal laptops.
Besides, it becomes extremely easy to work from anywhere and not 
to worry about working on the latest version of the code.

Parallel development on different parts of our project has been
possible and highly beneficial on account of the MVC model we
employed. Using CVS allowed us to quickly combine each other's
code updates of various parts. This greatly reduced the typical
use of email and other means even for sharing temporary trial codes.

\section{Conclusions}
A robust web application should employ the same principles of 
developing quality softwares. Only the tools and frameworks differ
between the two. Techniques will always largely depend upon what
kind of commitment the developers can put in the effort, as 
it will different for fulltime developers working in a company
and students with a finite number of semesters in school.

There are many scopes of improvement over our design, each with 
its pitfalls, but a good initial design can take care of most issues.
AJAX can be used from the beginning as it is increasingly becoming
the standard in producing web applications that behave like a 
standalone application. Its few pitfalls can be easily overcome
with a carefully designed workflow from the scratch.

With the advent of new frameworks, making a robust web application
that would run for a few years without getting too much outdated
is a challenging issue. Existing frameworks for making large scale
web application are not only huge but needs a lot of training to 
maintain and add features. A basic framework using java beans and
servlets is still a good solution for a medium scale application
like this and at the cost of less sophistication, it is rather
easy to be maintained by students.

\newpage
\section*{Appendix A}
  
  \subsection*{Usage of tools}
  The usage of the custom tools to generate code is given below:\\

  \parindent 0in
  {\bf BeanHelper}\\
  Usage: {\tt java BeanHelper table\_a field\_aa field\_bb ... field\_nn}
  
  This will generate the following code fragments:
  \begin{verbatim}
  -------------- SETTERS -------------
  b.setFieldAa(rs.getString("field_aa"));
  ...
  b.setFieldBb(rs.getString("field_nn"));
  -------------- ADDERS -------------
  String insertStmt = "INSERT INTO table_a(field_aa, ..., field_nn) VALUES ( xxxID.nextval,
  + "'" + b.getFieldAa() + "', "
  ...
  + "'" + b.getFieldNn() + "')";
  -------------- UPDATERS -------------
  String updateStmt = "UPDATE table_a SET "
  + "field_aa='" + b.getFieldAa() + "', "
  ...
  + "field_Nn='" + b.getFieldNn() + "' "
  + "WHERE field_nn='" + b.getFieldNn() + "'";
  -------------- TESTER -------------
  TableAManager = TableAManager.instance();
  TableABean b = new TableABean();
  b.setFieldAa("field_aa1");
  ...
  b.setFieldNn("field_nn1");
  -------------- HELPER for servlets ---------
  String fieldAa = request.getParameter("fieldAa");
  ...
  String fieldNn = request.getParameter("fieldNn");
  b.setFieldAa(fieldAa);
  ...
  b.setFieldNn(fieldNn);

  \end{verbatim}
  
  {\bf TableConv}\\
  Usage: {\tt java TableConv <spec\_file>}

  The spec\_file contains table conversion specification as follows:
  The first line contains source and destination table names
  separated by a whitespace. The following lines contain
  one source field and corresponding destination field separated
  by a whitespace, till a blank like. Multiple such specifications
  can be put in a single specification file. An example is:

  \begin{verbatim}
  language ps_su_apply_lnguag
  language_id SEQNUM
  user_id SU_APPLY_USER_ID
  language_name LANGUAGE_CD
  speak READ_PROFICIENCY
  read WRITE_PROFICIENCY
  write SPEAK_PROFICIENCY
  \end{verbatim}
  
\end{document}